\documentclass[aps,twocolumn,pra,tightenlines,superscriptaddress,footinbib,longbibliography]{revtex4-2}
\usepackage{amssymb}
\usepackage{amsmath}
\usepackage{graphicx}
\usepackage{txfonts}
\usepackage{color}
\usepackage{amsfonts}
\usepackage{bm}
\usepackage{mathrsfs}
\usepackage[colorlinks,linkcolor=blue,anchorcolor=blue,citecolor=blue,urlcolor=blue]{hyperref}
\usepackage[resetlabels]{multibib}
\usepackage{multirow}
\usepackage{makecell}
\usepackage{booktabs}
\usepackage{colortbl}
\definecolor{tabcolor}{rgb}{.6902,.18824,.37647}

\begin{document}
\setcounter{equation}{0} 
\setcounter{figure}{0}

\title{Quantum squeezing induced nonreciprocal phonon laser}

\author{Tian-Xiang Lu}
\affiliation{College of Physics and Electronic Information, Gannan Normal University, Ganzhou 341000, Jiangxi, China}
\affiliation{Key Laboratory of Low-Dimensional Quantum Structures and Quantum Control of Ministry of Education, Department of Physics and Synergetic Innovation Center for Quantum Effects and Applications, Hunan Normal University, Changsha 410081, China}
\author{Yan Wang}
\affiliation{Academy for Quantum Science and Technology, Zhengzhou University of Light Industry, Zhengzhou 450002, China}
\author{Keyu Xia}
\affiliation{College of Engineering and Applied Sciences, National Laboratory of Solid State Microstructures, and Collaborative Innovation\\ Center of Advanced Microstructures, Nanjing University, Nanjing 210023, China}
\author{Xing Xiao}
\affiliation{College of Physics and Electronic Information, Gannan Normal University, Ganzhou 341000, Jiangxi, China}
\author{Le-Man Kuang}
\affiliation{Key Laboratory of Low-Dimensional Quantum Structures and Quantum Control of Ministry of Education, Department of Physics and Synergetic Innovation Center for Quantum Effects and Applications, Hunan Normal University, Changsha 410081, China}
\affiliation{Academy for Quantum Science and Technology, Zhengzhou University of Light Industry, Zhengzhou 450002, China}
\author{Hui Jing}\email{jinghui73@gmail.com}
\affiliation{Key Laboratory of Low-Dimensional Quantum Structures and Quantum Control of Ministry of Education, Department of Physics and Synergetic Innovation Center for Quantum Effects and Applications, Hunan Normal University, Changsha 410081, China}
\affiliation{Academy for Quantum Science and Technology, Zhengzhou University of Light Industry, Zhengzhou 450002, China}

\date{\today}

\begin{abstract}
Phonon lasers or coherent amplifications of mechanical oscillations have provided powerful tools for both fundamental studies of coherent acoustics and diverse applications ranging from ultrasensitive force sensing to phononic information processing. Here, we propose how to achieve directional phonon lasing with an optomechanical resonator coupled to a nonlinear optical resonator. We find that, by pumping the nonlinear resonator, directional optical squeezing can occur along the pump direction. As a result, we can achieve the directional mechanical gain by utilizing the directional optical squeezing, thus leading to nonreciprocal phonon lasing with a well-tunable directional power threshold. Our work shows a feasible way to build nonreciprocal phonon lasers with various nonlinear optical mediums, which are important for such a wide range of applications as directional acoustic amplifiers, invisible sound sensing or imaging, and one-way phononic networks.
\end{abstract}
\date{\today}
\maketitle

\section{Introduction} \label{intro}
Phonon lasers or coherent amplifications of mechanical oscillations have played an essential role in the fundamental studies of coherent acoustics and applications ranging from sound imaging or sensing~\cite{application1,application2,application3,SP_PL,Force_sensing_using_phonon_laser} to topological motion control~\cite{application4} and phononic engineering~\cite{application5,application6}. In recent years, phonon lasers have been achieved in diverse platforms, such as trapped ions~\cite{ions_and_cold_atoms,PhysRevLett.131.043605}, cold atoms~\cite{ultra-cold}, optical tweezers~\cite{levitatedPL,tweezerPL1}, and cavity optomechanical (COM) systems~\cite{COM_PL1,COM_PL2,COM_PL3,PhysRevLett.127.073601,Wu_Haibin_PhysRevLett.124.053604,Wu_Haibin_PNAS}, to name only a few~\cite{ElectromechanicalPL,Superlattice1,Superlattice2,QuantumDotsPL1,QuantumDotsPL2,doi:10.1126/sciadv.adg7841}.
In particular, COM devices~\cite{COMreview}, suitable for integration on chip-scales, have been used to realize different kinds of phonon lasers~\cite{PT_PL1,PT_PL2,EP_PL,Phase_PL1,Phase_PL2,vector_PL1,vector_PL2,Dynamical_PL1,Dynamical_PL2}, such as exceptional-point phonon lasers~\cite{PT_PL1,PT_PL2,EP_PL}, phase-modulated phonon lasers~\cite{Phase_PL1,Phase_PL2}, and vector phonon lasers~\cite{vector_PL1,vector_PL2}. 

Very recently, nonreciprocal phononic devices, featuring directional flow of phonons, were explored and utilized for chiral phonon transport or cooling~\cite{Nonreciprocal_acoustic_nonlinear5,Nonreciprocal_acoustic_nonlinear6,lai2020nonreciprocal}, phonon isolation~\cite{fleury2014sound,Nonreciprocal_acoustic_nonlinear,zhang2015giant,Nonreciprocal_acoustic_nonlinear1,Nonreciprocal_acoustic_nonlinear3,Nonreciprocal_acoustic_nonlinear4}, one-way mechanical networks~\cite{Nonreciprocal_acoustic_PT,Nonreciprocal_acoustic_nonlinear2,Nonreciprocal_acoustic_metamaterials,wang2022acoustic,penelet2021broadband,Nonreciprocal_acoustic_metamaterials1,Nonreciprocal_acoustic_metamaterials2}, and backscattering-immune acoustic sensing or imaging~\cite{news_Nonreciprocal_acoustic,news_Sound_Isolation}. In these works, nonreciprocal acoustic control has been enabled by incorporating nonlinear mediums~\cite{Nonreciprocal_acoustic_nonlinear,zhang2015giant,Nonreciprocal_acoustic_nonlinear1,Nonreciprocal_acoustic_nonlinear3,Nonreciprocal_acoustic_nonlinear4,Nonreciprocal_acoustic_PT}, circulating fluids~\cite{fleury2014sound}, macroscopic metamaterials~\cite{Nonreciprocal_acoustic_nonlinear2,Nonreciprocal_acoustic_metamaterials,wang2022acoustic,penelet2021broadband}, space-time modulations~\cite{Nonreciprocal_acoustic_time1,Nonreciprocal_acoustic_time2}, and complicated synthetic structures~\cite{Nonreciprocal_acoustic_metamaterials2}. In particular, nonreciprocal phonon lasers were proposed by using the relativistic Sagnac effect in a spinning COM device or magnomechanical system~\cite{Nonreciprocal_PL1,Nonreciprocal_PL2}, making it possible to operate phonon lasers in a highly asymmetric way. However, in these works~\cite{Nonreciprocal_PL1,Nonreciprocal_PL2}, high-speed rotation of resonators is required and thus strongly relies on stable couplings between spinning devices and flying fibers. It is therefore highly desirable to seek a new approach that is free of any spinning component and easily tunable in optical ways. 

In a very recent work, directional quantum squeezing was proposed to achieve an optical diode or circulator~\cite{tang2022quantum}, opening up an attractive route to achieve nonreciprocal devices via harnessing various quantum resources. We note that the advantages of quantum squeezing have also been confirmed, e.g., significant amplifications of light-motion interactions or coherent couplings of hybrid quantum systems~\cite{PhysRevLett.114.093602,PhysRevA.100.062501,PhysRevLett.120.093601,Zhao_Wensensing,PhysRevA.99.023833,PhysRevLett.124.073602,PhysRevA.101.053826,PhysRevA.102.032601,PhysRevLett.127.093602,PhysRevLett.126.023602,villiers2023dynamically,PhysRevLett.122.030501,burd2021quantum,PhysRevLett.125.153602,lemonde2016enhanced,PhysRevApplied.17.024009,PhysRevA.107.023722,PhysRevLett.130.073602,WangYan_SCMPA,jiao2023tripartite}. Based on one-way squeezing, nonreciprocal photonic or magnonic devices have also been explored~\cite{PhysRevA.108.023716,Squeezing-induced_nonreciprocal_PB,10.1063/5.0158334,Huang:22magnonlaser}. Inspired by these pioneering works, here we study how to achieve a nonreciprocal phonon laser in a compound COM system by using directional optical squeezing. We find that by unidirectionally driving the nonlinear optical resonator, asymmetric coupling of the resonators can emerge, resulting in nonreciprocal mechanical gains and well-tunable direction-dependent threshold of phonon lasing. Also, we find that in such nonreciprocal devices, the mechanical gain exhibits squeezing-enhanced robustness against optical decays. Our scheme, requiring only two-mode optical-frequency matching and free of any spinning device, is promising to be realized under current experimental conditions and thus can be utilized for, e.g., nonreciprocal force sensing and chiral acoustic information processing or networking~\cite{Nonreciprocal_acoustic_nonlinear5,Nonreciprocal_acoustic_nonlinear6}. 

\begin{figure}[tbp]
\center
\includegraphics[width=0.95\columnwidth]{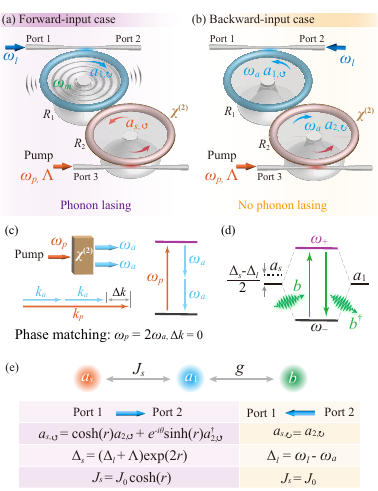}	\caption{(Color online) Operating a nonreciprocal phonon laser with directional squeezing. A strong pump with frequency $\omega_p$ and strength $\Lambda$, coming from port 3, causes the CCW mode to be squeezed, i.e., $a_{2,\circlearrowleft}\rightarrow a_{s,\circlearrowleft}$, in the nonlinear resonator $R_2$. (a) The input laser with frequency $\omega_l$ driven from port 1 excites the CW mode $a_{1,\circlearrowright}$ in $R_1$, which is coupled to the squeezed mode $a_{s,\circlearrowleft}$ in $R_2$ with an effective strength $J_s$, while as shown in (b), when driven from port 2, it excites the CCW mode $a_{1,\circlearrowleft}$ in $R_1$, coupled to the non-squeezed mode $a_{2,\circlearrowright}$in $R_2$ with the strength $J_0$. (c) The nonlinear parametric process: a pump photon with frequency $\omega_p$ and wave-number $k_p$ can be down-converted into two photons with the same frequency $\omega_a$ and wave-number $k_a$ due to the $\chi^{(2)}$ nonlinearity under the phase matching condition. (d) The energy-level diagram shows the mechanism of two-level phonon laser, i.e., stimulated emissions of phonons under the condition of population inversion (for the optical supermodes $\omega_\pm$). (e) The mode couplings of this system, with a comparison of direction-dependent key symbols. Here $r$ is the squeezing parameter.}
\label{fig1}
\end{figure}

\section{Theoretical model} \label{sec2}
As shown in Fig.~\ref{fig1}, we consider a compound COM system consisting of two coupled whispering-gallery-mode (WGM) microtoroid resonators and two nearby optical waveguides. The resonator $R_1$ of resonance frequency $\omega_a$ and decay rate $\kappa_1$ supports a mechanical breathing mode (with frequency $\omega_m$ and effective mass $m$), and is driven by a signal field of frequency $\omega_l$ from port 1 (or port 2) corresponding to the forward-input case (or backward-input case). The other resonator $R_2$ of resonance frequency $\omega_a$ and decay rate $\kappa_2$, manufactured of high-quality thin film with $\chi^{(2)}$-nonlinearity~\cite{PhysRevLett.92.043903,PhysRevLett.104.153901,PhysRevLett.117.123902,PhysRevApplied.6.014002,cite-keynonlinear,Bruch:19lithium,PhysRevLett.125.263602,cite-keygeneration,PhysRevLett.126.133601,cite-keythin-film,Lu:21niobate,Lu:20Lithium}, and thus supporting the parametric amplification process~\cite{PhysRevLett.114.093602}, is pumped from port 3 by a continuous coherent laser field with frequency $\omega_p$, strength $\Lambda$, and phase $\theta$. As shown in Fig.~\ref{fig1}(c), because of the directional phase-matching condition in the parametric nonlinear process~\cite{tang2022quantum,PhysRevLett.126.133601}, that is, the conservation of energy and momentum ($\omega_p=2\omega_a$, $\Delta k=0$), the counterclockwise (CCW) mode $a_{2,\circlearrowleft}$ is squeezed to a mode $a_{s,\circlearrowleft}$, but the clockwise (CW) mode $a_{2,\circlearrowright}$ is unsqueezed~\cite{tang2022quantum,PhysRevA.108.023716,Huang:22magnonlaser,Squeezing-induced_nonreciprocal_PB}. For the forward-input case, in a frame rotating at frequency $\omega_l$, the total Hamiltonian of this system can be written at the simplest level as ($\hbar=1 $):
\begin{align}\label{Eq1}
\mathcal{H} =&-\Delta_{l}a_{1,\circlearrowright}^{\dag}a_{1,\circlearrowright}-\Delta_{l}a_{2,\circlearrowleft}^{\dag}a_{2,\circlearrowleft}+\omega_{m}b^{\dag}b,\nonumber\\
 & + J_0(a_{1,\circlearrowright}^{\dag}a_{2,\circlearrowleft}+a_{2,\circlearrowleft}^{\dag}a_{1,\circlearrowright})-gx_{0}a_{1,\circlearrowright}^{\dag}a_{1,\circlearrowright}(b+b^{\dag}), \nonumber\\
&+i\varepsilon_{l}(a_{1,\circlearrowright}^{\dag}-a_{1,\circlearrowright}) +\frac{\Lambda}{2}  (a_{2,\circlearrowleft}^{\dagger 2} e^{-i \theta}+a_{2,\circlearrowleft}^{2} e^{i \theta}),
\end{align}
where $a_{1,\circlearrowright}$ ($a_{1,\circlearrowright}^{\dag}$), $a_{2,\circlearrowleft}$ ($a_{2,\circlearrowleft}^{\dag}$), and $b$ ($b^{\dag}$) are the annihilation (creation) operators of the CW mode in $R_1$, the CCW mode in $R_2$, and the mechanical mode in $R_1$, respectively. $\Delta_{l}=\omega_{l}-\omega_{a}$, $J_0$ (or $g=\omega_a/r_1$) denotes the coupling strength between the resonators (or the COM coupling strength), $x_0=\sqrt{\hbar/2m\omega_m}$, and $r_1$ is the radius of the COM resonator. $\varepsilon_{l} =\sqrt{2\kappa_1P_{\textrm{in}}/\hbar\omega_l}$ is the driving amplitude with input power $P_{\textrm{in}}$. To diagonalize $\mathcal{H}$, we define the squeezed operator $a_{s,\circlearrowleft}$ via the Bogoliubov transformation~\cite{PhysRevLett.114.093602,PhysRevLett.120.093601}:
\begin{align}a_{s,\circlearrowleft}=\cosh(r)a_{2,\circlearrowleft}+e^{-i\theta}\sinh(r)a_{2,\circlearrowleft}^{\dagger}~,\end{align}with the squeezing parameter $r=(1/4)\ln[(\Delta_l-\Lambda)/(\Delta_l+\Lambda)]$, which requires $\left|\Delta_l\right|>\left| \Lambda \right|$ to avoid the
system instability. Then, with the rotating wave approximation (see Appendix A for more details), the Hamiltonian of the system becomes
\begin{align}\label{Eq2}
\mathcal{H}_{\mathrm{f}}&=-\Delta_{l}a_{1,\circlearrowright}^{\dag}a_{1,\circlearrowright}-\Delta_{s}a_{s,\circlearrowleft}^{\dag}a_{s,\circlearrowleft}+J_s(a_{1,\circlearrowright}^{\dag}a_{s,\circlearrowleft}+a_{s,\circlearrowleft}^{\dag}a_{1,\circlearrowright}) \nonumber\\
&+\omega_{m}b^{\dag}b-gx_{0}a_{1,\circlearrowright}^{\dag}a_{1,\circlearrowright}(b+b^{\dag}) + i\varepsilon_{l}(a_{1,\circlearrowright}^{\dag}-a_{1,\circlearrowright}),
\end{align}
where \begin{align}\Delta_{s}=(\Delta_l+\Lambda)\exp(2r),~~J_s = J_0\cosh (r).\end{align}It is clearly seen that the effective squeezed mode detuning $\Delta_s$ and the effective coupling rate $J_s$ are modulated by the squeezing strength $\Lambda$ under the condition of ensuring the stability of the system (e.g., $\left|\Delta_l\right|>\left| \Lambda \right|$). For the backward-input case, the Hamiltonian of the system reads
\begin{align}\label{Eq3}
\mathcal{H}_{\mathrm{b}}=&-\Delta_{l}a^{\dag}_{1,\circlearrowleft}a_{1,\circlearrowleft}-\Delta_{l}a^{\dag}_{2,\circlearrowright}a_{2,\circlearrowright}+J_0(a^{\dag}_{1,\circlearrowleft}a_{2,\circlearrowright}+a_{2,\circlearrowright}^{\dag}a_{1,\circlearrowleft})\nonumber\\
 &+\omega_{m}b^{\dag}b-gx_{0}a_{1,\circlearrowleft}^{\dag}a_{1,\circlearrowleft}(b+b^{\dag}) + i\varepsilon_{l}(a_{1,\circlearrowleft}^{\dag}-a_{1,\circlearrowleft}).
 \end{align}
Comparing the Hamiltonians $\mathcal{H}_{\mathrm{f}}$ and $\mathcal{H}_{\mathrm{b}}$, it can be clearly seen that the detuning and coupling strengths of these two Hamiltonians are completely different due to the directional squeezing effect.

\begin{figure}[tbp]
\center
\includegraphics[width=0.98\columnwidth]{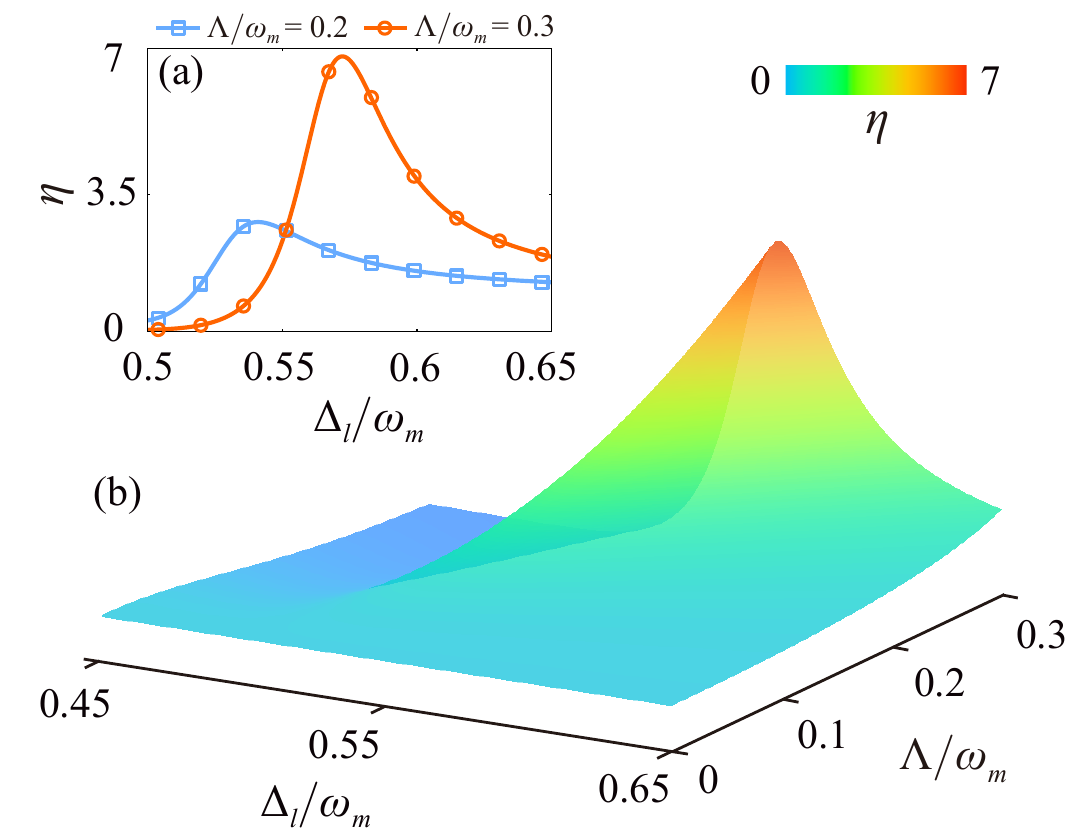}	
\caption{(Color online) (a) The mechanical displacement amplification factor $\eta$ as a function of the optical detuning $\Delta_{l}$, for different values of squeezing strength $\Lambda$. (b) $\eta$ versus $\Delta_{l}$ and $\Lambda$, with the experimentally accessible parameter values $J_0/\omega_m=0.5$ and $P_{\textrm{in}}=10\,\mu\mathrm{W}$.}
\label{fig2}
\end{figure}

Below, we show that for our COM system, this directional squeezing effect leads to distinct changes in the radiation pressure on the mechanical mode, hence resulting in a nonreciprocal phonon lasing action. Then, for the forward-input case, the steady-state solutions can be obtained as (see Appendix A for more details)
\begin{align}\label{Eq4}
\alpha_{1,\circlearrowright} &  =\frac{\varepsilon_{l}(\kappa_{2}-i\Delta_{s})}{(\kappa_{1}-i\Delta_{l}-igx_s)(\kappa_{2}-i\Delta_{s})+J_{s}^2}, \nonumber\\
\alpha_{s,\circlearrowleft}&  =\frac{J_{s}\alpha_{1,\circlearrowright}}{\Delta_{s}+i\kappa_{2}}, ~~~\beta=\frac{gx_s\vert \alpha_{1,\circlearrowright}\vert ^{2}}{\omega_{m}-i\gamma_{m}},
\end{align}
where $\gamma_{m}$ is the damping rate of the mechanical mode. In our calculations, for the backward-input case, we need to, respectively, replace $a_{1,\circlearrowright}$, $a_{s,\circlearrowleft}$, $J_{s}$, $\Delta_{s}$ with $a_{1,\circlearrowleft}$, $a_{2,\circlearrowright}$, $J_{0}$, $\Delta_{l}$. $x_s = x_0\,(\beta + \beta^{*} )$ is the steady-state mechanical displacement, and is proportional to
\begin{align}
x_{s}=\frac{\hbar g\vert \alpha_{1,\circlearrowright}\vert ^{2}}{m(\omega_m^2+\gamma_m^2)}.\end{align}
To see the effect of the squeezing effect on the radiation pressure of the mechanical mode, we defined the mechanical displacement amplification factor
\begin{align}\label{Eq5}
\eta=\frac{x_s(\Lambda \neq0)}{x_s(\Lambda =0)},
\end{align}
here $x_s(\Lambda \neq0)$ [$x_s(\Lambda=0)$] is the displacement corresponding to the forward-input case (backward-input case). Clearly, besides the driving field $\varepsilon_{l}$, the squeezing strength $\Lambda$ also affects the values of the mechanical displacement $x_s$. For the forward-input case, both the effective coupling rate $J_s$ and the effective squeezed mode detuning $\Delta_s$, can be adjusted by the squeezing strength $\Lambda$. Figures~\ref{fig2}(a) and \ref{fig2}(b) show the mechanical displacement amplification factor $\eta$ as a function of the optical detuning $\Delta_{l}$ and the squeezing strength $\Lambda$. It is seen that the mechanical displacement amplification factor can be adjusted due to the squeezing effect, that is, the steady-state mechanical displacement $x_s$ can be enhanced for the forward-input case. The amplified displacement indicates an enhancement of phonon generation~\cite{Nonreciprocal_PL1,Nonreciprocal_PL2}.

\begin{figure*}[htb]
	\center
	\includegraphics[width=1.9\columnwidth]{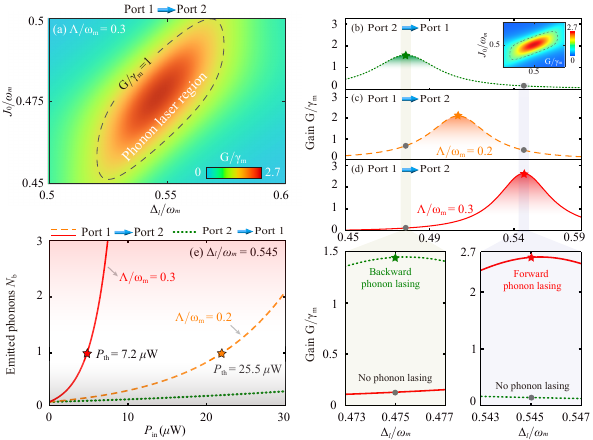} \caption{(Color online) (a) The mechanical gain $G$ versus the detuning $\Delta_l$ and the coupling strength $J_0$ for the forward-input case (port 1$\rightarrow$ port 2). (b-d) $G$ versus $\Delta_l$ for different input directions. The right inset shows $G$ versus $\Delta_l$ and $J_0$ for the backward-input case (port 2$\rightarrow$ port 1). Orange dashed and red solid curves correspond to the forward-input case for the squeezing strengths $\Lambda/\omega_m=0.2$ and $\Lambda/\omega_m=0.3$, respectively. The green dashed curve corresponds to the backward-input case with $\Lambda/\omega_m=0.2$ and $\Lambda/\omega_m=0.3$. (e) The stimulated emitted phonon number $N_b$ as a function of the pump power $P_{\textrm{in}}$ for different input directions. The five-pointed stars correspond to the threshold power $P_\mathrm{th}$, which is determined by the threshold condition $G=\gamma_m$. We choose $P_{\textrm{in}}=10\,\mu\mathrm{W}$ in (a-d), $J_0/\omega_m=0.48$ in (b-d), and $\Delta_l/\omega_m=0.545$ in (e).}
	\label{fig3}
\end{figure*}

Now we will show that nonreciprocal phonon lasing action can be achieved in such a system. By use of the supermode operators~\cite{COM_PL1,COM_PL2,COM_PL3}
\begin{align}
a_{\pm}=(a_{1,\circlearrowright}\pm a_{s,\circlearrowleft})/\sqrt{2},
\end{align}
the Hamiltonian of Eq.~(\ref{Eq2}) can be written as
\begin{align}\label{Eq6}
H=&~\omega_+a^\dag_+a_++\omega_-a^\dag_-a_-+\frac{\Delta}{2}(a^{\dagger}_{+}a_{-}+a^{\dagger}_{-}a_{+})\nonumber\\
&+\omega_mb^\dag b-\frac{g x_0}{2}(a^{\dagger}_{+}a_{-}b+b^{\dagger}a^{\dagger}_{-}a_{+})\nonumber\\
&+\frac{i\varepsilon_l}{\sqrt{2}}[(a^\dag_++a^\dag_-)-(a_++a_-)],
\end{align}
after applying the rotating-wave approximation~\cite{COM_PL1,Nonreciprocal_PL1,Nonreciprocal_PL2}, where effective mode frequencies
\begin{align}\omega_{\pm}=-(\Delta_s+\Delta_l)/2\pm J_s,
\end{align}
and $\Delta=\Delta_s-\Delta_l$. For the backward-input case, we have $\Delta=0$ and $J_s=J_0$. We note that, compared with the traditional phonon laser system (absorption and emission of phonons are described by the fourth term)~\cite{COM_PL1,COM_PL2,COM_PL3}, Eq.~(\ref{Eq6}) contains an additional detuning term (the third term). It implies that the coupling between the optical supermodes depends on the squeezing effect when driving the resonator from port 1. Thus, the phonon lasing process can be dramatically modified. Usually, the supermode operators are defined as $a_{\pm}=(a_{1,\circlearrowright}\pm a_{s,\circlearrowleft})/\sqrt{2}$ for coupled cavities with the same resonant frequency~\cite{COM_PL1,COM_PL2,Nonreciprocal_PL1,Nonreciprocal_PL2,Huang:22magnonlaser}. As mentioned earlier, the squeezing strength $\Lambda$ has brought some changes to our system, such as the effective squeezed mode detuning $\Delta_s$ and the effective coupling rate $J_s$. According to Ref.~\cite{tang2022quantum}, $J_s$ is enhanced exponentially relative to $J_0$ with increasing $\Lambda$. However, in our work, to ensure that this transformation can safely be simplified as we use the above operators [derived from Eq.~(\ref{Eq2}) to Eq.~(\ref{Eq6})], we do not consider the system to achieve strong coupling. Hence, the maximum value of squeezing strength is $0.3\omega_m$ in this work. For $\Lambda/\omega_m\leq0.3$, we have confirmed that the detuning $\Delta=\Delta_s-\Delta_l$ and the enhanced coupling rate $\Delta J=J_s-J_0$ are much smaller than $\Delta_l$ and $J_0$. For example, when $\Lambda/\omega_m=0.3$,
the detuning $\Delta/\omega_m$ is about 0.1, and the enhanced coupling rate $J_s/J_0$ is about 1.05. These mean that our calculations are reasonable and valid.

By using the ladder operator and population-inversion operator of the optical supermodes as~\cite{COM_PL1,COM_PL2}
\begin{align}
p=a^\dag_- a_+,~~~\delta n=a^\dag_+ a_+-a^\dag_-a_-,
\end{align}
respectively, the equations of motion of the system then become
\begin{align}\label{Eq7}
\dot{b}=&-(\gamma_m+i\omega_m)b+\frac{ig x_0}{2}p,\nonumber\\
\dot{p}=&-2(\kappa_0 +iJ_s)p+\frac{i}{2}(\Delta-g x_0b)\delta n +\frac{\varepsilon_l}{\sqrt{2}}(a_{+}+a^{\dagger}_{-}),
\end{align}
where $\kappa_0=(\kappa_1+\kappa_2)/2$. By using the standard procedures (see Appendix~B for more details), the mechanical gain $G$ can be obtained:
\begin{align}\label{Eq88}
G=&\,G_0+\mathcal{G}=\frac{g^2 x_0^2\kappa_0\delta n}{2(2J_s-\omega_m)^2+8\kappa_0^2} \nonumber\\
&\,\,+\frac{\varepsilon_l^2g^2 x_0^2(\omega_m-2J_s)(\Delta_s+\Delta_l)\kappa_0}{\left[4(2J_s-\omega_m)^2+16\kappa_0^2\right]\left[ D^2+(\Delta_l+\Delta_s)^2\kappa_0^2\right]},
\end{align}
with 
\begin{align}
D&=J_s^2+\kappa_0^2-\Delta_s\Delta_l+\frac{[g^2 x_0^2n_b-2gx_0\Delta\mathrm{Re}(b)]}{4},\nonumber \\ 
\delta n
&=\frac{\varepsilon_l^2\left[2J_s\Delta_{s}-\kappa_0 g x_0\mathrm{Im}(b)-J_s g x_0\mathrm{Re}(b)\right]}{D^2+\kappa_0^2(\Delta_l+\Delta_{s})^2},
\end{align}
where $n_b=b^{\dagger}b$ is the phonon number. We note that the first term $G_0$ in Eq.~(\ref{Eq88}) is proportional to the population inversion $\delta n$, and $\delta n$ depends on $\Delta_s$, which is quite different from the conventional phonon laser system without directional quantum squeezing effect. Also, the second term $\mathcal{G}$ in Eq.~(\ref{Eq88}) depends on $\Delta_s$. This indicates that different mechanical gains can be obtained for the forward-input and backward-input cases, which makes it possible to achieve a nonreciprocal phonon laser.

\section{Nonreciprocal phonon laser} \label{sec3}
In numerical simulations, to demonstrate that the observation of the phonon laser process is within current experimental reach, we have selected experimentally feasible parameters~\cite{COM_PL1}, i.e., $\omega_{a}=1.93\times10^5\,\mathrm{GHz} $, the optical quality factor $Q_1=9.7\times10^7$, $Q_2=4\times10^7$, $\kappa_{1,2}=\omega_{a}/Q_{1,2}$, $2r_1=66\,\mu\mathrm{m}$, $2r_2=69\,\mu\mathrm{m}$, $\omega_{m}/2\pi=23.4\,\mathrm{MHz} $, $m=50\,\mathrm{ng}$, $\gamma_m=0.24$\,MHz, and $\Lambda/\omega_m=0.3$, and thus $\Delta/\omega_m\sim0.1$ and $J_s/J_0\sim1.05$.

In Fig.~\ref{fig3}(a), the calculated mechanical gain $G$ is plotted as a function of the optical detuning $\Delta_l$ and the coupling strength $J_0$ for the forward-input case. It is clearly shown that, for $\Lambda/\omega_m=0.3$, the mechanical gain $G/\gamma_m>1$ can be obtained with the proper selection of $\Delta_l$ and $J_0$, making phonon lasing possible. A maximum mechanical gain of $G/\gamma_m\simeq2.7$ is obtained with $\Delta_l/\omega_m\sim0.545$ and $J_0/\omega_m\sim0.48$. This means that the strongest phonon lasering occurs due to the resonance of the input drive field with the supermode $\omega_+$ (i.e. driving the up energy level in Fig.~\ref{fig1}(d))~\cite{COM_PL1,COM_PL2,COM_PL3}. In Figs.~\ref{fig3}(b-d), we plot the dependence of mechanical gain $G$ on input directions of the driving field. For the backward-input case, the maximum is located around $\Delta_l/\omega_m\sim0.475$, corresponding to a backward phonon laser [see Fig.~\ref{fig3}(a)]. However, for the forward-input case, the squeezing effect leads to a blueshift in the maximum of the mechanical gain $G$ [see the orange dashed curve in Fig.~\ref{fig3}(c) and the red solid curve in Fig.~\ref{fig3}(d)]. That is, for $\Lambda/\omega_m=0.3$, by driving the system from its left side (port 1$\rightarrow$ port 2), we can get maximum mechanical gain at $\Delta_l/\omega_m\sim0.545$ [see Fig.~\ref{fig3}(c)], corresponding to a forward phonon laser.

\begin{figure}[tbp]
\center
\includegraphics[width=0.98\columnwidth]{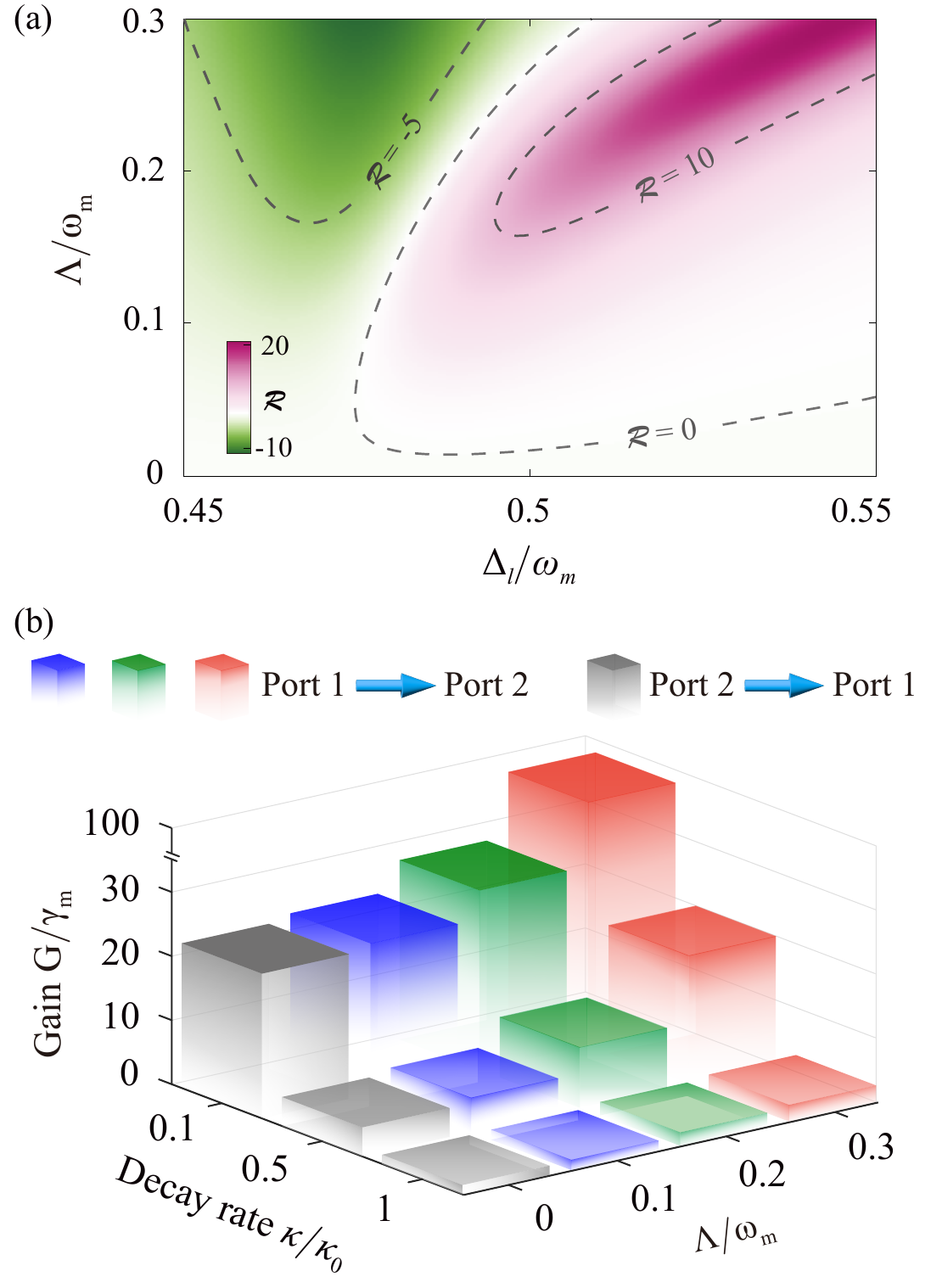}	
\caption{(Color online) (a) Dependence of the isolation parameter $\mathcal{R}$ on the detuning $\Delta_l$ and the squeezing strength $\Lambda$. (b) The optimal mechanical gain $G$ versus the scaled decay rate $\kappa/\kappa_0$ and the squeezing strength $\Lambda$ for different input directions, with the experimentally accessible parameter values $J_0/\omega_m=0.48$ and $P_{\textrm{in}}=10\,\mu\mathrm{W}$.}
\label{fig4}
\end{figure}

The underlying physical mechanism can be explained as follows: In the forward-input case, the squeezing effect causes a frequency shift to $a_{s,\circlearrowleft}$ with respect to the bare mode $a_{2,\circlearrowleft}$~\cite{tang2022quantum}, leading to $G/\gamma_m>1$ around $\Delta_l/\omega_m\sim0.545$. In contrast, in the backward-input case, the squeezing effect does not contribute anything to the $a_{2,\circlearrowright}$, corresponding to a conventional phonon laser ($G/\gamma_m>1$ around $\Delta_l/\omega_m\sim0.475$ with $J_0/\omega_m=0.48$)~\cite{COM_PL1,COM_PL2,COM_PL3}. This implies that, by adjusting the detuning $\Delta_l$ and squeezing strength $\Lambda$, one can get enhanced or significantly suppressed mechanical gain by driving the device from the left or right side, or vice versa. Thus, our system provides a new route to realize a nonreciprocal phonon laser by changing the input directions. Compared with the method of using a rotating cavity to realize a nonreciprocal phonon laser~\cite{Nonreciprocal_PL1,Nonreciprocal_PL2}, our scheme only needs two-mode matching in one resonator, without its high-speed rotation, and is easier to realize experimentally. Thus, our work provides a new way to realize a nonreciprocal phonon laser.

According to Eq.\,(\ref{Eq88}), the stimulated emitted phonon number $N_b$ can be calculated; that is,
\begin{align}
N_b=\mathrm{exp}[2(G-\gamma_m)/\gamma_m],
\end{align}
which characterizes the performance of the phonon laser~\cite{COM_PL1}. Then, from the above expression and under the threshold condition of phonon laser $N_b=1$ (i.e., $G=\gamma_m$)~\cite{COM_PL1}, we can get the threshold pump power
\begin{align}
P_\mathrm{th}\approx\frac{2\hbar\kappa_0\gamma_m\omega_l [(J_s^2+\kappa_0^2-\Delta_l\Delta_s)^2+\kappa_0^2(\Delta_s+\Delta_l)^2]}{\kappa_1 J_sg^2x_0^2\Delta_s},
\end{align}
in which we have used $|b_s|^2\ll 1$ at the threshold. It is clearly seen that the directional quantum squeezing effect has a significant impact on the threshold power $P_{\textrm{th}}$. In Fig.\,\ref{fig3}(e), we show the stimulated emitted phonon number $N_b$ as a function of the pump power $P_{\textrm{in}}$. For the forward-input case (i.e., $\Lambda/\omega_m=0.3$), the threshold power is about $7.2\,\mu\mathrm{W}$ at a fixed $\Delta_l/\omega_m=0.545$ [corresponding to the maximal value of the mechanical gain in Fig.\,\ref{fig3}(d)], which approaches the threshold of about $7\,\mu\mathrm{W}$ reported in experiment~\cite{COM_PL1}. In comparison, for the backward-input case (without the squeezing effect), the mechanical gain is suppressed at $\Delta_l/\omega_m=0.545$, so a larger threshold power is required to implement the phonon laser.

To clearly see the squeezing effect on the nonreciprocal phonon lasing action, we introduce the isolation parameter
\begin{align}
\mathcal{R}=10\log_{10}\frac{N_b(\Lambda\neq0)}{N_b(\Lambda=0)},
\end{align}
where $N_b\,(\Lambda\neq0)$ [or $N_b\,(\Lambda=0)$] is the emitted phonon number corresponding to the forward-input case (or backward-input case). Figure \ref{fig4}(a) shows the isolation parameter $\mathcal{R}$ versus the optical detuning $\Delta_l$ and the squeezing strength $\Lambda$. A nonzero isolation parameter $\mathcal{R}$ indicates that nonreciprocity occurs in the phonon lasing action. It is found that when the detuning regions are selected correctly, nonreciprocity occurs. For example, for the detuning $\Delta_l/\omega_m\sim0.545$ and squeezing strength $\Lambda/\omega_m=0.3$, forward phonon lasing can be realized [see Figs.~\ref{fig3}(b) and \ref{fig3}(d)], which is an inevitable result from the difference between $\delta n\,(\Lambda\neq0)$ and $\delta n\,(\Lambda=0)$. So, in such a COM system, nonreciprocal phonon lasing can be realized by directional quantum squeezing.
\begin{table*}[t]
\footnotesize
\caption{Experimental parameters of the phonon laser in a coupled-microtoroid COM system}
\label{tab1}
\tabcolsep 9.5pt
\begin{tabular*}{\textwidth}{cccccccc}
\arrayrulecolor{tabcolor}
\toprule
\toprule
References&Resonators& Geometry&Diameter &Optical quality& Mechanical& Mechanical quality
& Power threshold\\
&(Material) & & $(\mu\mathrm{m})$& factors&frequency $(\mathrm{MHz})$&factors&($\,\mu\mathrm{W}$)\\
\toprule
\cite{COM_PL1}& $R_{1}$~(Si)&Microtoroid &$\sim66\,\mu\mathrm{m}$& $9.7\times10^7$&$147.1 $&$1\times10^3$ &$7$ \\
&$R_{2}$~(Si)&Microtoroid &$\sim69\,\mu\mathrm{m}$& $3\times10^7$&-&-&\\
\toprule[0.3pt]
\cite{COM_PL2}&$R_{1}$~(Si)&Microtoroid
&$\sim68\,\mu\mathrm{m}$& $9.25\times10^7$&$59.2$
&$18\times10^3$ &1.2 \\
&$R_{2}$~(Si)&Microtoroid&$\sim69\,\mu\mathrm{m}$& $2.5\times10^7$&-&-&\\
\toprule[0.3pt]
\cite{COM_PL3}&$R_{1}$~(Si)&Microtoroid
&$\sim34.5\,\mu\mathrm{m}$& $6.33\times10^7$&$17.38$
&$0.4\times10^3$ &2.5 \\
&$R_{2}$~(Si)&Microtoroid&$\sim34\,\mu\mathrm{m}$& $1.5\times10^7$&-&-&\\
\bottomrule
\bottomrule
\end{tabular*}
\caption{Performances of various $\chi^{(2)}$-nonlinearity integrated mediums. PPLN: periodically poled $\mathrm{LiNbO}_3$}
\label{tab2}
\tabcolsep 7.5pt 
\begin{tabular*}{\textwidth}{cccccccc}
\arrayrulecolor{tabcolor}
\toprule
\toprule
 References& Material &$\chi^{(2)}$~($\mathrm{pm/v}$)&Geometry  &Size ($\mu\mathrm{m}$): rings: $r/w/h$,
&Optical quality&Nonlinear single-photon&Pump\\
& & & & disks:$r/h$, and spheres:$r$& factor&coupling strength~($\mathrm{MHz}$)&power~($\mathrm{mW}$)\\
\toprule[0.5pt]
\cite{PhysRevLett.92.043903}&PPLN&0.023&Microdisk &$1500/500$&$2.0\times10^8$&-&$25$ \\
\cite{PhysRevLett.104.153901}&$\mathrm{LiNbO}_3$&-& Microtoroid&$1900/500$&$3.4\times10^7$&-&$0.03$\\
\cite{PhysRevLett.117.123902}&AlN&1.3&Microring&30/1.12/1.0&$1.8\times10^5$&0.74 &$24.4$\\
\cite{PhysRevApplied.6.014002}&$\mathrm{LiNbO}_3$&2.1&Microdisk&51/0.7&$1.1\times10^5$&-&$10$\\
\cite{cite-keynonlinear}&$\mathrm{SiO}_2$&-&Microsphere &62&$4.8\times10^7$&-&$0.9$\\
\cite{Bruch:19lithium}&AlN&6&Microring&60/1.2/1.0 &$1.0\times10^6$&0.5&$11$\\
\cite{PhysRevLett.125.263602}&AlN&27&Microring&55/1.6/2&$2.6\times10^5$&-&$0.35$\\
\cite{cite-keygeneration}&$\mathrm{Si}_3\mathrm{N}_4$&0.2&Microring&23/1.2/0.6&$1.2\times10^6$&-&$15$\\
\cite{PhysRevLett.126.133601}&$\mathrm{Si}_3\mathrm{N}_4$&-&Microring&23/1.2/1.1 &$1.38\times10^6$&0.25
&$22.4$\\
\cite{cite-keythin-film}&$\mathrm{LiNbO}_3$&0.6&Microring&80/1.6/0.6&$9.0\times10^5$&-&$20$\\
\cite{Lu:21niobate}&PPLN&40&Microring&-&$6.0\times10^5$&7.4&$0.03$\\
\cite{Lu:20Lithium}&PPLN&-&Microring&75/-/-&$1.8\times10^6$&1.2&$0.1$\\
\bottomrule
\end{tabular*}
\end{table*}

In experiments, the decay rate of the photon modes can be engineered, for example, by placing an external nanotip near a microresonator~\cite{peng2014loss}. The effect of the normalized decay rate of photon mode $\kappa/\kappa_0$ on the optimal mechanical gain $G$ for different input directions is shown in~\ref{fig4}(b). It can be found that for the same value of squeezing strength $\Lambda/\omega_m$, the the optimal mechanical gain $G$ of the forward-input case can reach a much higher value than that of the opposite direction due to the directional quantum squeezing, indicating that the forward phonon laser tends to become more robust against decay rates of photon modes.

\section{Experimental feasibility}
Phonon lasers have been demonstrated experimentally in a wide range of physical systems~\cite{tweezerPL1,ions_and_cold_atoms,ultra-cold,ElectromechanicalPL,Superlattice1,Superlattice2,QuantumDotsPL1,QuantumDotsPL2,levitatedPL,COM_PL1,COM_PL2,COM_PL3,PhysRevLett.127.073601,Wu_Haibin_PhysRevLett.124.053604,Wu_Haibin_PNAS}. In particular, optically pumped phonon lasing has been demonstrated in a compound COM system consisting of two coupled silica microtoroid WGM resonators and a nearby optical fiber~\cite{COM_PL1,COM_PL2,COM_PL3}. As already confirmed in an experiment~\cite{COM_PL1}, the first resonator $R_1$ (with a diameter of $r_1=66\,\mu\mathrm{m}$) supports a high-$Q$ ($Q_1=9\times10^7$) optical WGM mode and a mechanical mode with mechanical quality factor $Q_m=1\times10^3$ and resonance frequency $\omega_m=2\pi\times23.4\,\mathrm{MHz}$. The second resonator $R_2$ (with a diameter of $r_2=69\,\mu\mathrm{m}$) supports a pure optical WGM mode with optical quality factor $Q_2=3\times10^7$. By placing a tapered optical fiber near the resonator, a tunable laser can be coupled into $R_1$ via the optical evanescent field. Also, the two resonators can be coupled through the evanescent field with the coupling strength $J_0$, which can be adjusted by controlling the air gap between the resonators. For a controllable gap $0.2\,\mu\mathrm{m}\sim 2\,\mu\mathrm{m}$, $J_0$ is typically between $5\,\mathrm{MHz}\sim5\,\mathrm{GHz}$~\cite{COM_PL1}. In such a COM system, when the upper state (supermode $\omega_+$) is occupied by a sufficient number of photons coming from the driving laser through the nearby optical waveguide, these photons begin to split into the lower frequency photons of the lower state (supermode $\omega_-$) and the coherent phonons, i.e., a phonon amplifier and phonon laser with a power threshold of about $P_{\textrm{th}}=7\,\mu\mathrm{w}$~\cite{COM_PL1}. As shown in Table~\ref{tab1}, we compare the performance of recent phonon laser experiments based on a composite COM system consisting of two coupled silica microtoroid WGM resonators and a nearby optical waveguide~\cite{COM_PL1,COM_PL2,COM_PL3}.

Recently, in experimental manufacturing~\cite{PhysRevLett.92.043903,PhysRevLett.104.153901,PhysRevLett.117.123902,PhysRevApplied.6.014002,cite-keynonlinear,Bruch:19lithium,PhysRevLett.125.263602,cite-keygeneration,PhysRevLett.126.133601,cite-keythin-film,Lu:21niobate,Lu:20Lithium}, microring resonators with large $\chi^{(2)}$-nonlinearity and high-$Q$ can be fabricated using various thin film materials, such as silicon nitride, lithium niobate, and aluminum nitride. Table~\ref{tab2} shows more relevant parameters for experimentally achieving a $\chi^{(2)}$-nonlinear resonator. We note that the $Q$ factors of lithium-niobate-based resonators can reach $10^7$ or $10^8$~\cite{PhysRevLett.92.043903,PhysRevLett.104.153901,PhysRevApplied.6.014002}, which are better suited for our study. For an experimentally feasible quality factor $Q\simeq4\times10^7$, the decay rate of a resonator with frequency $\omega_{a}=1.93\times10^5\,\mathrm{GHz}$ is $\kappa\simeq48.2\,\mathrm{MHz}$. According to Eq.~(\ref{EqA7}) (see Appendix~A), the relationship between the power $P_p$ of a pump field and the squeezing strength $\Lambda$ is given by~\cite{tang2022quantum,Huang:22magnonlaser} \begin{align}
\Lambda=\sqrt{8 g_d^2 \kappa_{2} P_p/\hbar \omega_p \kappa_p^2},
\end{align}
where $g_d$ denotes nonlinear single-photon coupling strength in the parametric nonlinear process, and $\kappa_{p}$ denotes the external decay rate for the pump field. By choosing experimentally feasible values~\cite{PhysRevLett.92.043903,PhysRevLett.104.153901,PhysRevLett.117.123902,PhysRevApplied.6.014002,cite-keynonlinear,Bruch:19lithium,PhysRevLett.125.263602,cite-keygeneration,PhysRevLett.126.133601,cite-keythin-film,Lu:21niobate,Lu:20Lithium}: $g_d=0.025\,\mathrm{MHz}$, $\kappa_{2}=48.2\,\mathrm{MHz}$, $\kappa_p=2\kappa_2$, and $\omega_p=2\omega_a$. we choose pump power $P_p=0.02\,\mu\mathrm{W}$ or $P_p=30\,\mathrm{mW}$ leading to $\Lambda/\omega_m\sim0.2$ or $\Lambda/\omega_m\sim290$. Therefore, we strongly believe that our proposed scheme is experimentally feasible.
\section{Conclusion} \label{sec4}
In conclusion, we have studied a nonreciprocal phonon laser in a compound COM system consisting of an optomechanical resonator and a $\chi^{(2)}$-nonlinear resonator. By unidirectionally pumping the nonlinear resonator, the squeezed effect occurs only in the selected direction, which significantly modifies the mechanical gain and power threshold, resulting in the nonreciprocal phonon lasing. Moreover, we find that in such nonreciprocal devices, the mechanical gain exhibits squeezing-enhanced robustness against optical decays. Our results opens up a new route to manipulate COM devices by using the directional quantum squeezing, and may find intriguing applications in designing phonon chips or high-precision motion sensors. Our scheme also opens many possibilities for further research: studying the role of directional quantum squeezing in enhancing or steering, for example, phonon blockade, macroscopic entanglement~\cite{PhysRevLett.125.143605}, and backscattering-immune force sensing~\cite{Force_sensing_using_phonon_laser}, in which quantum noise terms should be included.

\begin{acknowledgments}
H.J. is supported by the National Natural Science Foundation of China (NSFC, Grant No. 11935006), Hunan provincial major sci-tech program (2023ZJ1010), and the Science and Technology Innovation Program of Hunan Province (Grant No. 2020RC4047). L.-M.K. is supported by the NSFC (Grants No. 1217050862, 11935006 and 11775075). K.X. is supported by he National Key R$\&$D Program of China (Grants No. 2019YFA0308704), the National Natural Science Foundation of China (Grants No. 92365107), and the Program for Innovative Talents and Teams in Jiangsu (Grant No. JSSCTD202138). T.-X.L. is supported by the NSFC (Grant No. 12205054), the Jiangxi Provincial Education Office Natural Science Fund Project (GJJ211437), and Ph.D. Research Foundation (BSJJ202122). X.X. is supported by the NSFC (Grant No. 12265004). Y.W. is supported by the NSFC (Grant No. 12205256), and the Henan Provincial Science and Technology Research Project (Grant No. 232102221001).
\end{acknowledgments}
~
\section*{{APPENDIX A: Derivation of the Hamiltonian}} \label{appendix A}
We consider a compound COM system of two coupled resonators with the same resonance frequency $\omega_a$. One of the resonators $R_1$ supports a mechanical breathing mode with frequency $\omega_m$ and effective mass $m$. The second resonator $R_2$ is a purely optical resonator that is coupled to the first resonator via an evanescent field with a coupling strength of $J_0$. For simplicity, all directional subscripts are omitted. In the case of forward-input, the total Hamiltonian of this compound COM system can be written as ($\hbar=1 $):
\begin{align}\label{EqA1}
H^{\prime}   =&~\omega_{a}a_{1}^{\dag}a_{1}+\omega_{a}a_{2}^{\dag}a_{2}+\omega_{m}b^{\dag}b+ J_0(a_{1}^{\dag}a_{2}+a_{2}^{\dag}a_{1})\nonumber\\
&+\omega_{c}c^{\dag}c+g_d(a_{2}^{\dag2}c+a_{2}^{2}c^{\dag})-gx_{0}a_{1}^{\dag}a_{1}(b+b^{\dag})\nonumber \\
&+ i\varepsilon_{l}(a_{1}^{\dag}e^{-i\omega_lt}-a_{1}e^{i\omega_lt}) +i\lambda_{p}(c^{\dag}e^{-i\omega_pt}-ce^{i\omega_pt}),
\end{align}
where $\omega_{c}$ is the frequency of the second-harmonic modes in $R_2$. $g$ and $g_d$ are the COM coupling rate in the radiation-pressure process and the nonlinear single-photon coupling strength in the parametric nonlinear process. $\varepsilon_{l} =\sqrt{2\kappa_1P_{\textrm{in}}/\hbar\omega_l}$ is the driving amplitude with input power
$P_{\textrm{in}}$, $\lambda_{p} =\sqrt{2\kappa_2P_{p}/\hbar\omega_{p}}$ is the pump light with the power $P_{p}$, here $\kappa_1$ and $\kappa_2$ are the decay rates. By using the unitary transformation $U=\exp \left[\left(-i \frac{\omega_p}{2} a_1^{\dagger} a_1-i \frac{\omega_p}{2} a_2^{\dagger} a_2-i \omega_p c^{\dagger} c\right) t\right]$, the Hamiltonian $H^{\prime}$ can be transformed into the rotating frame, i.e.,
\begin{align}
H^{\prime\prime}=U^{\dagger}H^{\prime}U-iU^{\dagger}\frac{\partial U}{\partial t}.
\end{align}
Then we have
\begin{align}\label{EqA2}
H^{\prime\prime}   =&~-\Delta^{\prime}a_{1}^{\dag}a_{1}-\Delta^{\prime}a_{2}^{\dag}a_{2}+\omega_{m}b^{\dag}b+ J_0(a_{1}^{\dag}a_{2}+a_{2}^{\dag}a_{1})\nonumber\\
&-\Delta_{c}c^{\dag}c+g_d(a_{2}^{\dag2}c+a_{2}^{2}c^{\dag})-gx_{0}a_{1}^{\dag}a_{1}(b+b^{\dag}) \nonumber\\
&+ i\varepsilon_{l}(a_{1}^{\dag}e^{-i\Delta_{\textrm{in}}t}-a_{1}e^{i\Delta_{\textrm{in}}t}) +i\lambda_{p}(c^{\dag}-c),
\end{align}
where $\Delta^{\prime}=\omega_p/2-\omega_a$, $\Delta_{c}=\omega_p-\omega_c$, and $\Delta_{\textrm{in}}=\omega_l-\omega_p/2$. The dynamical equation of $c$ can be solved by the Heisenberg equation
\begin{align}\label{EqA3}
\dot{c}=\left(i \Delta_c-\kappa_p\right) c+\lambda_{p}-i g_d a_2^2.
\end{align}
Here, we consider the strong pump field to excite mode $c$ in $R_2$~\cite{tang2022quantum}. Then we get the steady-state solution
\begin{align}\label{EqA4}
c_s=\frac{-\lambda_p}{i \Delta_c-\kappa_p}.
\end{align}
After that, the Hamiltonian can be rewritten as
\begin{align}\label{EqA5}
H^{\prime\prime}   =&~-\Delta^{\prime}a_{1}^{\dag}a_{1}-\Delta^{\prime}a_{2}^{\dag}a_{2}+\omega_{m}b^{\dag}b\nonumber\\
&+ J_0(a_{1}^{\dag}a_{2}+a_{2}^{\dag}a_{1})-gx_{0}a_{1}^{\dag}a_{1}(b+b^{\dag})\nonumber \\
&+ i\varepsilon_{l}(a_{1}^{\dag}e^{-i\Delta_{\textrm{in}}t}-a_{1}e^{i\Delta_{\textrm{in}}t})+\frac{\Lambda}{2} (a_{2}^{\dagger 2} e^{-i \theta}+a_{2}^{2} e^{i \theta}),
\end{align}
where the strength and the phase are
\begin{equation}
\Lambda=2 g_d \sqrt{\frac{2 \kappa_{2} P_p}{\left(\Delta_c^{2}+\kappa_p^2\right) \hbar \omega_p}}, \quad \theta=-\operatorname{Arg}\left( c_{s}\right).
\end{equation}
Due to the resonance condition, we have $\omega_p=\omega_c$. Thus the pump power can be
obtained as
\begin{equation}\label{EqA7}
P_p=\frac{\hbar \omega_p \kappa_p^2 \Lambda^2}{8 g_d^2 \kappa_2}.
\end{equation}

In a frame rotating at frequency $\Delta_{\textrm{in}}$, the total Hamiltonian of this system can be written in the simplest level as
\begin{align}\label{EqA8}
\mathcal{H}   =&~-\Delta_la_{1}^{\dag}a_{1}-\Delta_la_{2}^{\dag}a_{2}+\omega_{m}b^{\dag}b\nonumber\\
&+ J_0(a_{1}^{\dag}a_{2}+a_{2}^{\dag}a_{1})-gx_{0}a_{1}^{\dag}a_{1}(b+b^{\dag}) \nonumber\\
&+ i\varepsilon_{l}(a_{1}^{\dag}-a_{1})+\frac{\Lambda }{2} (a_{2}^{\dagger 2} e^{-i \theta}+a_{2}^{2} e^{i \theta}),
\end{align}
where $\Delta_{l}=\Delta_{\textrm{in}}-\Delta^{\prime}=\omega_{l}-\omega_{1}=\omega_{l}-\omega_{2}$. To diagonalize $\mathcal{H}$, we define the squeezed operator $a_{s}$ via the Bogoliubov transformation~\cite{PhysRevLett.114.093602,PhysRevA.100.062501,PhysRevLett.120.093601} $a_{s}=\cosh (r)a_{2}+e^{-i\theta}\sinh (r)a_{2}^{\dagger}~,$ with the squeezing parameter $r=(1/4)\ln[(\Delta_l-\Lambda)/(\Delta_l+\Lambda)]$. The Hamiltonian $\mathcal{H}$ can be rewritten as
\begin{align}\label{EqA9}
\mathcal{H} =&-\Delta_la_{1}^{\dag}a_{1}+\omega_{m}b^{\dag}b-gx_{0}a_{1}^{\dag}a_{1}(b+b^{\dag})\nonumber\\
&-[\Delta_l(\cosh ^2 (r)+\sinh ^2 (r))+2\Lambda \cosh (r) \sinh (r)]a_{s}^{\dag}a_{s}\nonumber\\
&-\Delta_l\sinh ^2 (r) - \Lambda \cosh (r)\sinh (r)+ i\varepsilon_{l}(a_{1}^{\dag}-a_{1})\nonumber\\
&+ J_0\cosh (r)(a_{1}^{\dag}a_{s}+a_{s}^{\dag}a_{1})\nonumber\\&-J_0\sinh (r)(e^{-i\theta}a_{1}^{\dag}a_{s}^{\dag}+e^{i\theta}a_{1}a_{s}).
\end{align}
Then, with the rotating wave approximation and neglecting the constant term~\cite{PhysRevLett.114.093602,PhysRevA.100.062501,PhysRevLett.120.093601}, the Hamiltonian of the system can be changed into
\begin{align}\label{EqA10}
\mathcal{H}=&~-\Delta_{l}a_{1}^{\dag}a_{1}-\Delta_{s}a_{s}^{\dag}a_{s}+J_s(a_{1}^{\dag}a_{s}+a_{s}^{\dag}a_{1})\nonumber \\
&+\omega_{m}b^{\dag}b-gx_{0}a_{1}^{\dag}a_{1}(b+b^{\dag}) + i\varepsilon_{l}(a_{1}^{\dag}-a_{1}),
\end{align}
where $\Delta_{s}=(\Delta_l+\Lambda)\exp(2r),~~J_s = J_0\cosh (r).$ This Hamiltonian sets the stage for our calculations of the mechanical gain and the threshold power in the forward-input case. For the backward-input case, we have $J_s=J_0$, and $\Delta_s=\Delta_l$. Then, the Heisenberg equations of motion are written as
\begin{align}\label{Eq3}
\dot{a}_{1} &  =(i\Delta_l-\kappa_{1}){a}_{1}+igx_0(b+b^{\dag}){a}_{1}
-iJ_{s}{a}_{s} +\varepsilon_{l}, \nonumber\\
\dot{a}_{s} &  =\left(  i\Delta_{s}-\kappa_{2}\right){a}_{s}-iJ_{s}{a}_{1},\nonumber\\
\dot{b} &  =-\left(  i\omega_{m}+\gamma_{m}\right) b-igx_{0}{a}_{1}^{\dag}{a}_{1},
\end{align}
where $\gamma_{m}$ is the damping rate of the mechanical mode. As already confirmed in experiments with a COM-based phonon laser~\cite{COM_PL1,COM_PL2,COM_PL3}, for a strong driving field, the input noise terms can be safely ignored if one is interested only in the mean-number behaviors (i.e., the threshold feature of the mechanical gain or the phonon amplification). Then, the steady-state solutions can be obtained as
\begin{align}
\alpha_{1} &  =\frac{\varepsilon_{l}(\kappa_{2}-i\Delta_{s})}{(\kappa_{1}-i\Delta_{l}-igx_s)(\kappa_{2}-i\Delta_{s})+J_{s}^2}, \nonumber\\
\alpha_{s}&  =\frac{J_{s}\alpha_{1}}{\Delta_{s}+i\kappa_{2}}, ~~~\beta=\frac{gx_s\vert \alpha_{1}\vert ^{2}}{\omega_{m}-i\gamma_{m}},
\end{align}
where $x_s = x_0\,(\beta + \beta^{*} )$ is the steady-state mechanical displacement.

\section*{{APPENDIX B: Derivation of the mechanical gain}} \label{appendix B}
We introduce the supermode operators $a_{\pm}=(a_1\pm a_s)/\sqrt{2}$, which satisfy the commutation relations
$[a_{+}, a_{+}^\dag]=[a_{-}, a_{-}^\dag]=1,
[a_{+}, a_{-}^\dag]=0$.
Eq.\,(\ref{EqA10}) can be written as
\begin{align}\label{EqA23}
H=&~\omega_+a^\dag_+a_++\omega_-a^\dag_-a_-+\omega_mb^\dag b,\nonumber\\
&-\frac{g
x_0}{2}[(a_{+}^{\dagger}a_++a_{-}^{\dagger}a_-)+(a_+^{\dagger}a_-+a_-^{\dagger}a_+)](b^{\dagger}+b)\nonumber\\
&+\frac{\Delta}{2}(a^\dag_+a_-+a^\dag_-a_+)+\frac{i\varepsilon_l}{\sqrt{2}}[(a^\dag_++a^\dag_-)-(a_++a_-)],
\end{align}
with the supermode frequencies
$\omega_{\pm}=-(\Delta_s+\Delta_l)/2\pm J_s,$ and $\Delta=\Delta_s-\Delta_l$. Under the rotating-wave approximation condition $2J_s+\omega_m,\,\omega_m\gg|2J_s-\omega_m|$~\cite{Nonreciprocal_PL1,Nonreciprocal_PL2}. Thus we have
\begin{align}\label{EqA24}
H=&~\omega_+a^\dag_+a_++\omega_-a^\dag_-a_-+\frac{\Delta}{2}(a^{\dagger}_{+}a_{-}+a^{\dagger}_{-}a_{+})\nonumber\\
&+\omega_mb^\dag b-\frac{g x_0}{2}(a^{\dagger}_{+}a_{-}b+b^{\dagger}a^{\dagger}_{-}a_{+})\nonumber\\
&+\frac{i\varepsilon_l}{\sqrt{2}}[(a^\dag_++a^\dag_-)-(a_++a_-)].
\end{align}
In the supermode picture, the dynamical equations of the system can be written as
\begin{align}
\dot{a}_+&=-(i\omega_++\kappa_0)a_++\frac{i}{2}(g x_0b-\Delta)a_-+\frac{\varepsilon_l}{\sqrt{2}},\nonumber\\
\dot{a}_-&=-(i\omega_-+\kappa_0)a_-+\frac{i}{2}(g x_0b^\dag-\Delta)a_++\frac{\varepsilon_l}{\sqrt{2}},\nonumber\\
\dot{b}&=-(i\omega_m+\gamma_m)b+\frac{ig x_0}{2}a_+a^\dag_-,
\end{align}
where $\kappa_0=(\kappa_1+\kappa_2)/2$. With the ladder operator $p=a_{-}^\dag a_{+}$ and population inversion operator $\delta n=a_{+}^\dag a_{+}-a_{-}^\dag a_{-}$, the dynamical equations of the system are then read
\begin{align}\label{EqA25}
\dot{b}=&-(\gamma_m+i\omega_m)b+\frac{ig x_0}{2}p,\\
\dot{p}=&-2(\kappa_0 +iJ_s)p+\frac{i}{2}(\Delta-g x_0b)\delta n +\frac{\varepsilon_l}{\sqrt{2}}(a_{+}+a^{\dagger}_{-}).
\end{align}
By setting the time derivatives of $a_{\pm}$ and $p$ as zero, we can solve the steady-state solutions of the system, i.e.,
\begin{align}\label{EqA26}
p&=\frac{\sqrt{2}\varepsilon_l(a_++a^\dagger_-)-i(g x_0b-\Delta)\delta n}{2i(2J_s-\omega_m)+4\kappa_0},\nonumber\\
a_+&=\frac{\varepsilon_l(2i\omega_-+2\kappa_0+ig x_0b-i\Delta)}{2\sqrt{2}[D-i(\Delta_l+\Delta_s)\kappa_0]},\nonumber\\
a_-&=\frac{\varepsilon_l(2i\omega_++2\kappa_0+ig x_0b^\dag-i\Delta)}{2\sqrt{2}[D-i(\Delta_l+\Delta_s)\kappa_0]},
\end{align}
with
\begin{align}
D=J_s^2+\kappa_0^2-\Delta_s\Delta_l+\frac{[g^2 x_0^2n_b-2gx_0\Delta\mathrm{Re}(b)]}{4}, \nonumber
\end{align}
where $n_b=b^\dag b$. Substituting Eq.\,(\ref{EqA26}) into the dynamical equation of $b$ in Eq.\,(\ref{EqA25}) results in
\begin{align}
\dot{b}&=(-i\omega_m-i\omega'+G-\gamma_m)b+F,
\end{align}
where
\begin{align}
\omega'=&~\frac{g^2 x_0^2(2J_s-\omega_m)\delta n}{4(2J_s-\omega_m)^2+16\kappa_0^2}\nonumber\\
&+\frac{g^2 x_0^2\varepsilon_l^2 \kappa_0^2(\Delta_l+\Delta_s)}{[2(2J_s-\omega_m)^2+8\kappa_0^2][D^2+(\Delta_l+\Delta_s)^2\kappa_0^2]},\nonumber\\
F=&~\frac{g x_0\Delta \delta n}{4i(2J_s-\omega_m)+8\kappa_0}\nonumber\\
&+\frac{ig x_0\varepsilon_l^2[D(\kappa_0-iJ_s)+\kappa_0(\Delta_l+\Delta_s)\Delta_s]}{[2i(2J_s-\omega_m)+4\kappa_0][D^2+(\Delta_l+\Delta_s)^2\kappa_0^2]},\nonumber
\end{align}
and the mechanical gain is
\begin{align}
G=&\,G_0+\mathcal{G}\nonumber\\
=&~\frac{g^2 x_0^2\kappa_0\delta n}{2(2J_s-\omega_m)^2+8\kappa_0^2}\nonumber\\
&\,\,+\frac{\varepsilon_l^2g^2 x_0^2(\omega_m-2J_s)(\Delta_s+\Delta_l)\kappa_0}{\left[4(2J_s-\omega_m)^2+16\kappa_0^2\right]\left[ D^2+(\Delta_l+\Delta_s)^2\kappa_0^2\right]},
\end{align}
with
\begin{align}
\delta n
=\frac{\varepsilon_l^2\left[2J_s\Delta_{s}-\kappa_0 g x_0\mathrm{Im}(b)-J_s g x_0\mathrm{Re}(b)\right]}{D^2+\kappa_0^2(\Delta_l+\Delta_{s})^2}.\nonumber
\end{align}

\end{document}